\documentclass[english,jou]{apa}
\usepackage[T1]{fontenc}
\usepackage[latin9]{inputenc}
\usepackage{babel}
\usepackage{array}
\usepackage{url}
\usepackage{amsmath}
\usepackage{amssymb}
\usepackage{graphicx}
\usepackage[authoryear]{natbib}
\usepackage[unicode=true,pdfusetitle,
 bookmarks=true,bookmarksnumbered=false,bookmarksopen=false,
 breaklinks=true,pdfborder={0 0 1},backref=false,colorlinks=false]
 {hyperref}

\makeatletter

\providecommand{\tabularnewline}{\\}

\helvetica
\author{Author} 
\affiliation{Affiliation} 

\makeatother

\begin{document}

\title{Media Environment, Dual Process and Polarization: A Computational
Approach}

\author{In-Ho Yi}

\affiliation{New York University}

\abstract{News articles of varying degrees of truthfulness and political alignment,
and their influences on the political opinions of the media consumers
are modeled as a Bayesian network incorporating a mixture of ideas
from dual-reasoning models of Motivated Reasoning and Analytic/Intuitive
Reasoning. The result shows that as the media environment moves towards
the Post-Truth world, the problem of political polarization becomes
exacerbated.}
\maketitle

\section{Introduction}

``Fake News'' came to the spotlight in a recent surge of the use
of the term. It has been at least partially attributed to for recent
upsets in elections (\citet{BREXITTRUMP}), Anti-Vaxxers and other
movements based on false medical information (\citet{WASZAK2018115}),
and global warming denial (\citet{INNOCULATE}). Fake News is also
blamed for the radicalization of fringe movements and the rise of
racially and religiously motivated hate-crimes (\citet{antimuslim,hopenothate}).

In this introductory section, we first will set the scene by characterizing
the recent development of Fake News and the new media environment.
Then, we will cite studies that are of particular interest to the
present study. Finally, we will give an overview of the plan to model
and measure the effect of the media.

\subsection{The new media environment}

Two features of the new development are of interest to us. First is
the relativization of the concept of Truth. In the post-truth politics,
the line between assertions and emotional fleas and the truth is blurred.
With the advent of the new media environment, traditional journalists
are no longer seen as the gatekeepers of the fact-based public opinion,
and the notion of Truth has lost its privileged in the media landscape,
and we have seen an increase in media outlets that capitalizes on
various propaganda of questionable truthfulness. (\citet{doi:10.1177/2041905816680417},
\citet{DBLP:journals/corr/abs-1902-01752}).

Second is the polarization effect. It has long been observed that
the advent of the new media environment has the effect of accelerating
the polarization of public opinion (\citet{doi:10.1177/0266382117722446}).
From the ``One-Step flow'' point of view that the news readers exposure
to the media directly shape the news reader's own opinion (\citet{10.2307/25097863}),
as the political alignment of post-truth media approaches fringe politics,
readers of such news media become more polarized. On top of that,
various back-firing effects have been observed, whereby readers of
particular political mindset exercise Motivated Reasoning in discarding
the opposite side of the story and further reinforces the existing
belief.

These two characteristics of the new media have pushed the political
landscape to an alarming level of radicalization, as observed from
various election upsets to the rise of hate crime and other consequences
of the fringe movements. In this paper, we will model the different
media environments by allocating media consumption to premium centrist
media, premium partisan media and Fake News media. We will discuss
the media environment model in detail in the Methods section.

\subsection{Recent studies on political information consumption}

We list recent studies to draw inspiration from, to model the behavior
of the media consumers.

\citet{MOTPROCESS} show the effect of dis-confirmation bias of Motivated
Reasoning. The result suggests that, when processing political news,
readers tend not to exercise the same level of scrutiny to determine
the truthfulness of the news item.

\citet{PENNYCOOK2018} found that the ability to tell factually accurate
news from Fake News is positively correlated to the participants'
performance in the Cognitive Reflection Test. This result suggests
that Motivated Reasoning alone is not sufficient to account for the
discrepancies between individual media consumer's ability to single
out Fake News. 

Rather than completely replacing the existing observations on Motivated
Reasoning, we will use the result as a complementary force in determining
the probabilistic degree to which the agent will allow factual inaccuracies,
as we shall see in the following section.

\citet{Bail9216} found evidence for the backfiring effect, whereby
participants who were regularly exposed to the opposing views were
found to be even more polarized. In this study, to account for such
backfiring, we will count those that the agent will deem as Fake News
to have the opposite effect. This Fake News backfiring effect, combined
with the Motivated Reasoning phenomena, should be able to account
for the backfiring effect in general. We won't introduce a sub-module
for backfiring in general, however, as we believe that Motivated Reasoning
+ Fake News Backfiring should suffice to account for the backfiring
results in general, given the current media environment.

Now that we have examined a couple of recent empirical studies, we
will turn to the theoretical underpinning of the present study.

\subsection{Polarization and Bayesian network}

\citet{Jern2014} found that Bayesian networks that have root nodes
other than the hypothesis node can capture the polarization or convergence
effect, where two agents of diverging effects will either be polarized
or converge to the center after examining the same data point. As
we shall see later in this paper, the topology of the Bayesian network
presented in this paper has the ``truth judgment'' intervening node
that encompasses Motivated Reasoning and Fake News Backfiring. Unlike
\citet{Jern2014}, however, we use explicit programming constructs
available to us in a probabilistic programming language to model the
backfiring effect, and we do so on a continuous distribution. Nevertheless,
the basic idea that the intervening root node in a Bayesian network
can account for polarization remains intact.

\section{Methods}

In this section, we explain the design of the model and give the rationale
for design decisions. On the media side, we will first examine the
definition of a news item. Then we will discuss the abstraction of
a media outlet, and we will finally discuss the media environment.
On the agent side, we will discuss a Bayesian network in which the
truthfulness of a news item is determined by an agent.

\subsection{News item}

A news item has two numerical properties: \emph{politics} $\left(p_{n}\right)$
and \emph{truth} $\left(t_{n}\right)$.

\emph{$p_{n}$} value is centered at 0. The further away from the
center the politics value is, the more extreme the political alignment
of the news item is. If the absolute value is close to 1 or above,
the politics of the news item is considered pretty extreme.

\emph{$t_{n}$} value is 0 or above. Truth value of 1 or above is
considered factually accurate in every detail. Being true to the concept
of Post-Truth politics, \emph{$n_{t}$} is a continuous variable,
rather than a Boolean value.

\subsection{Media outlet}

Each type of media outlet emits a news item of probabilistic \emph{politics}
and \emph{truth} values. We define three different types of media
outlets and their \emph{$p_{n}$} and $t_{n}$ distributions.
\begin{center}
\begin{tabular}{|c|c|c|}
\hline 
Outlet & $p_{n}$ & \emph{$t_{n}$}\tabularnewline
\hline 
Premium Centrist & $\mathcal{N}\left(0,0.5\right)$ & $\mathcal{N}\left(0.8,0.2\right)$\tabularnewline
\hline 
Premium Partisan & $\mathcal{N}\left(\pm0.7,0.3\right)$ & $\mathcal{N}\left(0.8,0.2\right)$\tabularnewline
\hline 
Fake News Partisan & $\mathcal{N}\left(\pm0.9,0.1\right)$ & $\mathcal{N}\left(0.4,0.5\right)$\tabularnewline
\hline 
\end{tabular}
\par\end{center}

For partisan media outlets, their \emph{$p_{n}$} is bi-modal with
equal weight on both sides i.e. for Premium Partisan media:

\begin{align*}
\mathbb{P}\left(p_{n}\mid X=1\right) & \sim\mathcal{N}\left(0.7,0.3\right)\\
\mathbb{P}\left(p_{n}\mid X=0\right) & \sim\mathcal{N}\left(-0.7,0.3\right)\\
\mathbb{P}\left(X=0\right)= & \mathbb{P}\left(X=1\right)=\frac{1}{2}
\end{align*}

Similarly, Fake News Partisan media is also bi-modal, with more extreme
means.

\subsection{Media environment}

Media environment defines a mixture of media outlets. At each observation,
the media environment makes a weighted random choice among its mixture
of outlet, and let the chosen outlet to generate a news item. Here
are the media environments and their mixtures:
\begin{center}
\begin{tabular}{|>{\centering}p{0.2\columnwidth}|>{\centering}p{0.18\columnwidth}|>{\centering}p{0.2\columnwidth}|>{\centering}p{0.2\columnwidth}|}
\hline 
Media\\
Environment & Premium\linebreak{}
Centrist & Premium\linebreak{}
Partisan & Fake News\linebreak{}
Partisan\tabularnewline
\hline 
\hline 
ME1 & 70\% & 20\% & 10\%\tabularnewline
\hline 
ME2 & 40\% & 50\% & 10\%\tabularnewline
\hline 
ME3 & 30\% & 10\% & 60\%\tabularnewline
\hline 
\end{tabular}
\par\end{center}

ME1 represents an ideal media environment, with a majority of centrist
media, with a few partisan media outlets and very few Fake News media.
ME2 represents an environment where media outlets are more opinionated.
ME3 represents a post-truth media environment with an alarmingly large
amount of Fake News circulation. Such shift of media environments
from ME1 to ME3 over time can be attributed to the economic reality
for news media to find an audience in the increasingly crowded media
industry (\citet{BERNHARDT20081092,badnews}).

Now we turn to explain how agent processes news items. 

\subsection{Agent model}

An agent has two numerical properties: \emph{politics} $\left(p_{a}\right)$
and \emph{analytic} $\left(a_{a}\right)$. Following \citet{PENNYCOOK2018},
we assign the \emph{$a_{a}$} value as an intrinsic property of an
agent. While \emph{$p_{a}$} gets updates as the agent processes news
items, the \emph{analytic} value stays the same throughout the process.

Like news item, the prior distribution of \emph{$p_{a}$} of an agent
is centered around 0. The prior distribution is $p_{a}\sim\mathcal{N}\left(0,1\right)$.
\emph{$a_{a}$} value is sampled from $a_{a}\sim\text{Uniform}\left(0.5,1\right)$.

\subsection{Truth and politics judgment by agent}

Upon observing a news item, the agent makes a Boolean decision of
whether the news item is true or false. In order to make the decision,
the agent first calculates the motivational discount $d_{m}$:

\begin{align*}
d_{m} & =0.2\cdot0.2^{\left|p_{n}-p_{a}\right|}
\end{align*}

That is, the discount exponentially decays as the $p_{n}$ gets further
away from $p_{a}$. This reflects the observations of Motivated Reasoning
that agents are less likely to call out Fake News that are more aligned
to their own politics.

Two variables are sampled:
\begin{align*}
x_{n} & \sim\text{Uniform}\left(0,\min\left(0,t_{n}\right)\right)\\
x_{a} & \sim\text{Uniform}\left(0,\min\left(0,a_{a}-d_{m}\right)\right)
\end{align*}

\emph{truth judgment} $\left(t_{j}\in\left\{ \text{true},\text{false}\right\} \right)$
is true iff $x_{n}>x_{a}$:

\begin{align*}
\mathbb{P}\left(t_{j}=\text{true}\mid x_{n}>x_{a}\right) & =1\\
\mathbb{P}\left(t_{j}=\text{true}\mid x_{n}\le x_{a}\right) & =0
\end{align*}

Politics judgment $\left(p_{j}\right)$ reflects the agent's perception
of the political effect of the news item. When $t_{j}$ is false,
$p_{j}$ is the reverse of $p_{n}$.

\begin{align*}
p_{j} & =\begin{cases}
p_{n} & t_{j}\text{ is true}\\
-p_{n} & t_{j}\text{ is false}
\end{cases}
\end{align*}

This formulation reflects the backfiring effect of perceived Fake
News.

\subsection{Updating $p_{a}$}

$p_{j}$ is the observation value with which $p_{a}$ is updated.
The formulation of the Bayesian update is as following:

\begin{align*}
\mathbb{P}\left(p_{a}\mid p_{j}\right) & \propto\mathbb{P}\left(p_{a}\right)\mathbb{P}\left(p_{j}\mid p_{a}\right)
\end{align*}

where the likelihood function $\mathbb{P}\left(p_{j}\mid p_{a}\right)$
is the pdf of distribution $P\sim\mathcal{N}\left(p_{a},0.25\right)$

\begin{figure}
\begin{centering}
\includegraphics[width=0.9\columnwidth]{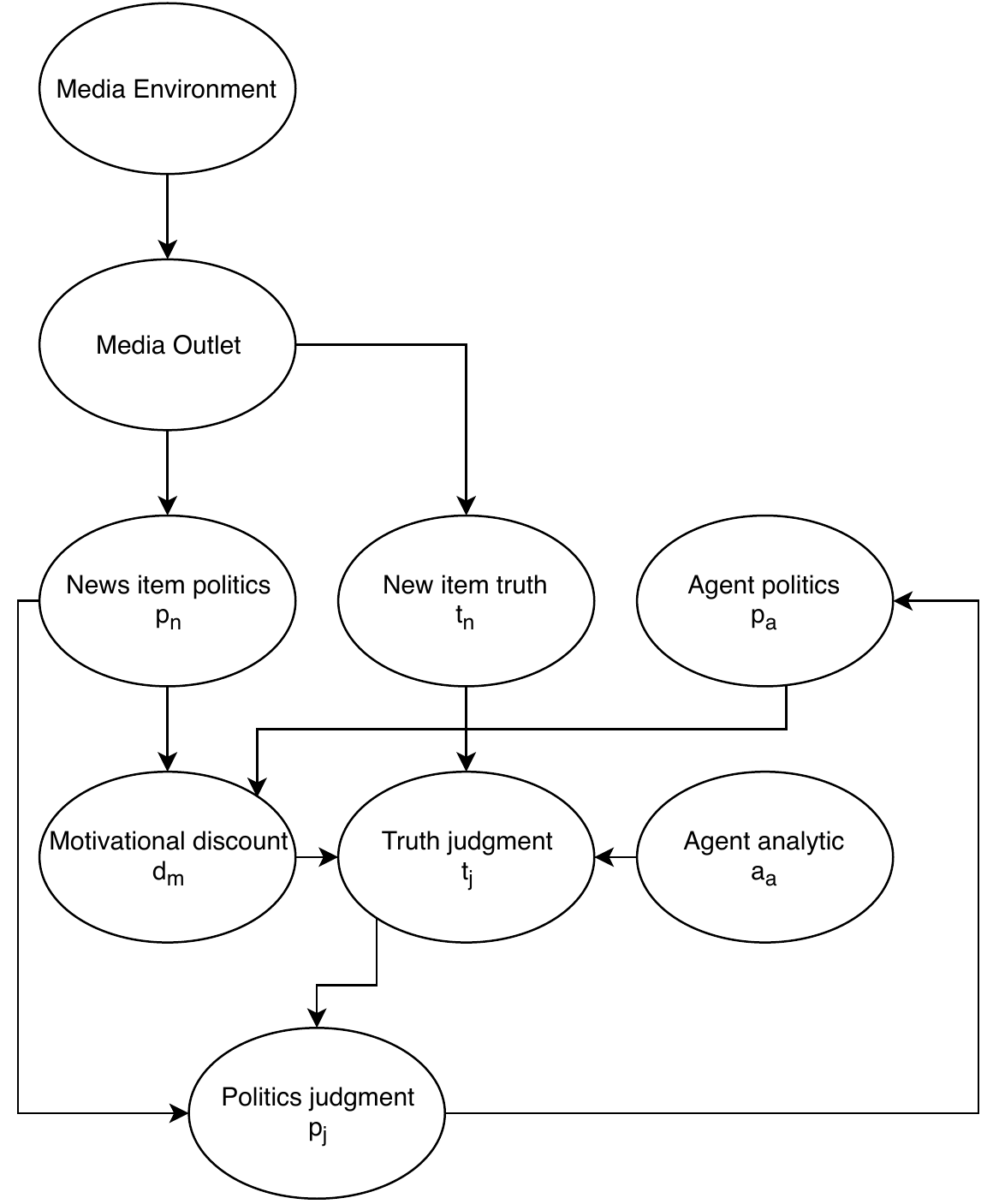}
\par\end{centering}
\caption{\label{fig:Bayesian-network-depicting}Bayesian network depicting
the model}

\end{figure}

In summary, Figure \ref{fig:Bayesian-network-depicting} shows the
model as a Bayesian network.

\subsection{Implementation of the model}

We used Anglican probabilistic programming language (\citet{progprgm})
to program the model. In order to obtain the resulting distribution,
we have run the model with Parallel Lighweight Metropolis-Hastings
algorithm with 5,000 particles, and obtained the first 200,000 iterations
and drew the histograms of the results\footnote{Source code for the model is available at \url{https://gitlab.com/chajath/polarize-fakenews-model}}.

\section{Results}

\begin{figure*}[p]
\begin{centering}
\includegraphics[width=0.3\textwidth]{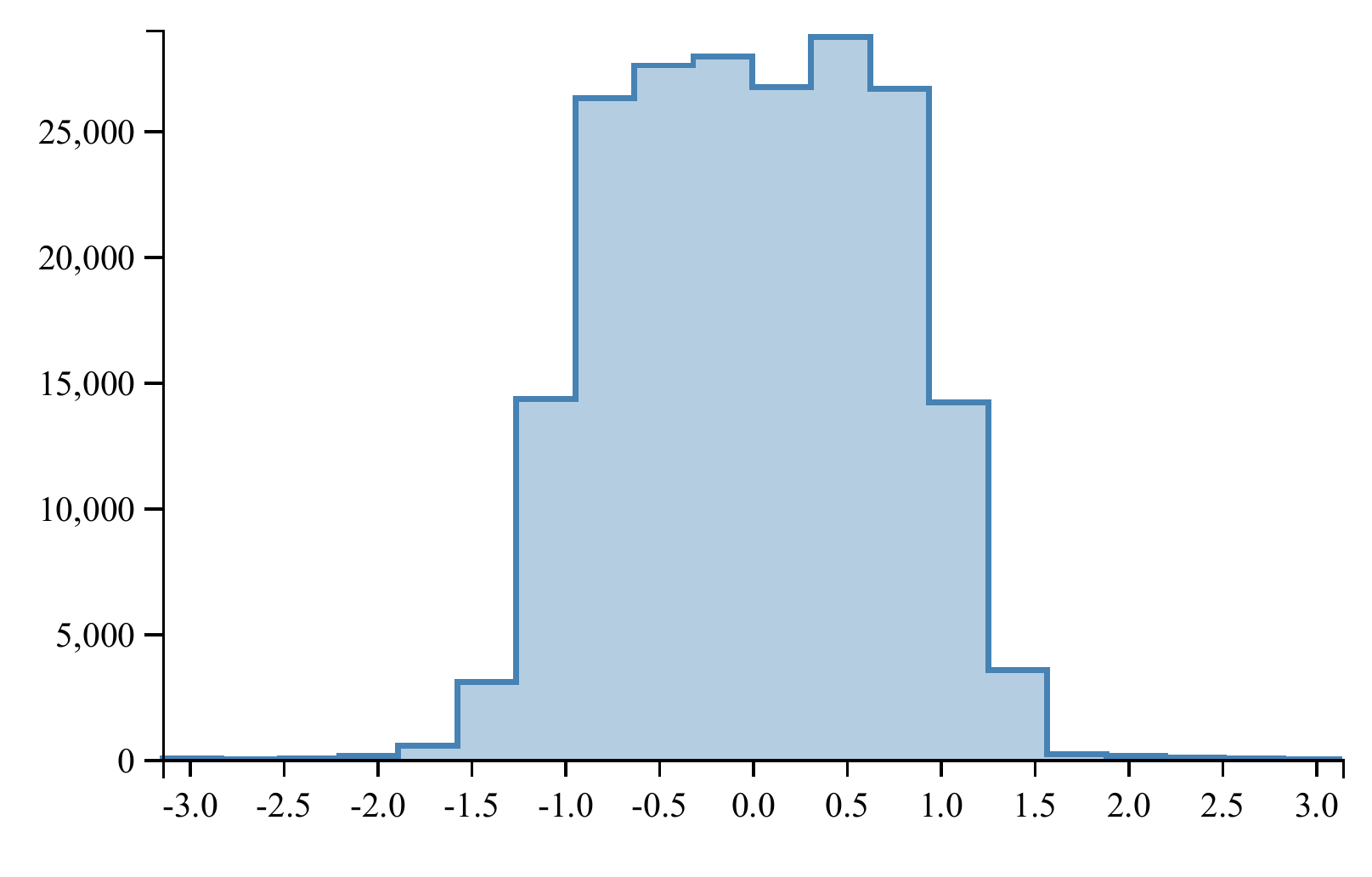}\includegraphics[width=0.3\textwidth]{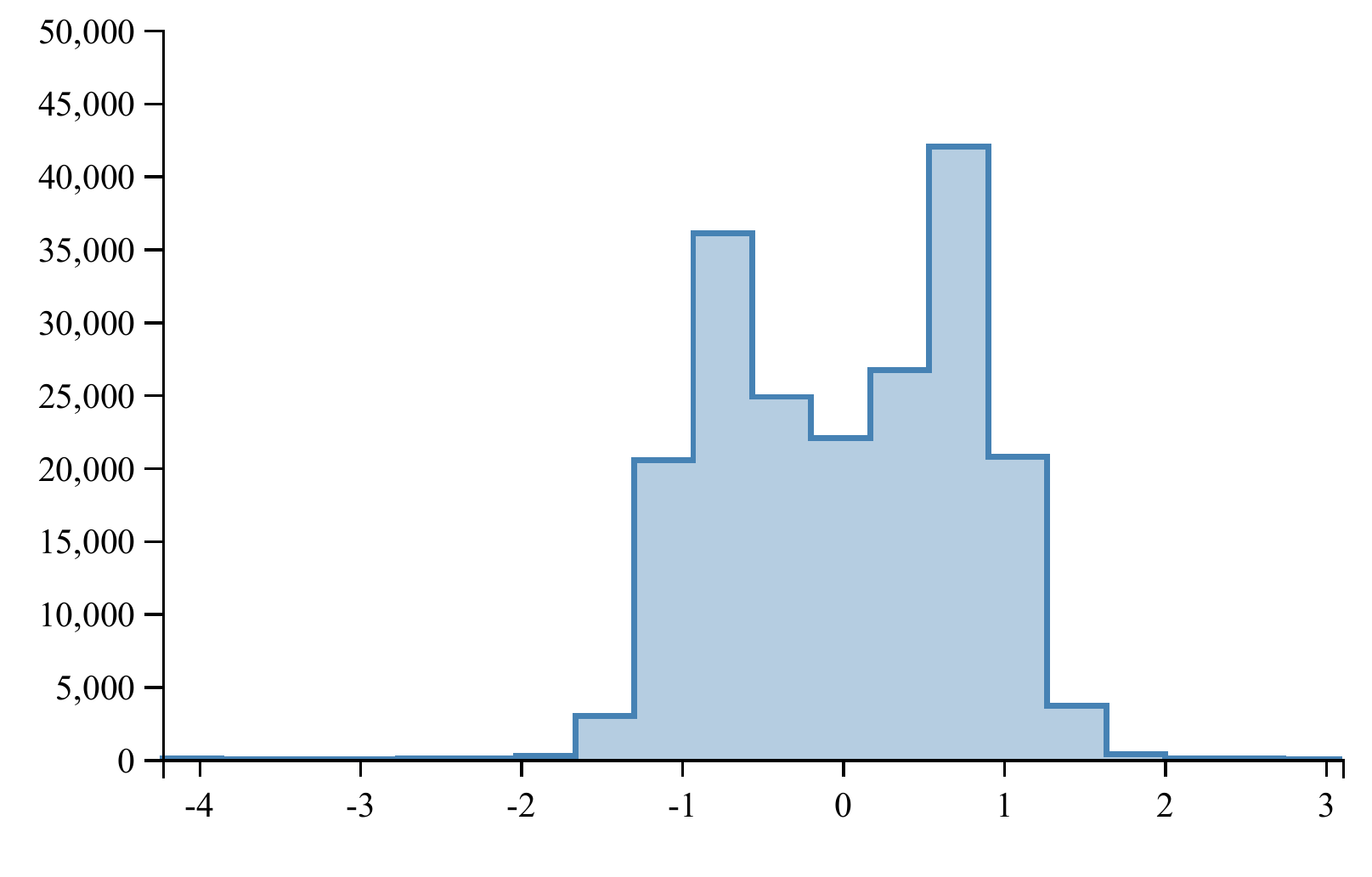}\includegraphics[width=0.3\textwidth]{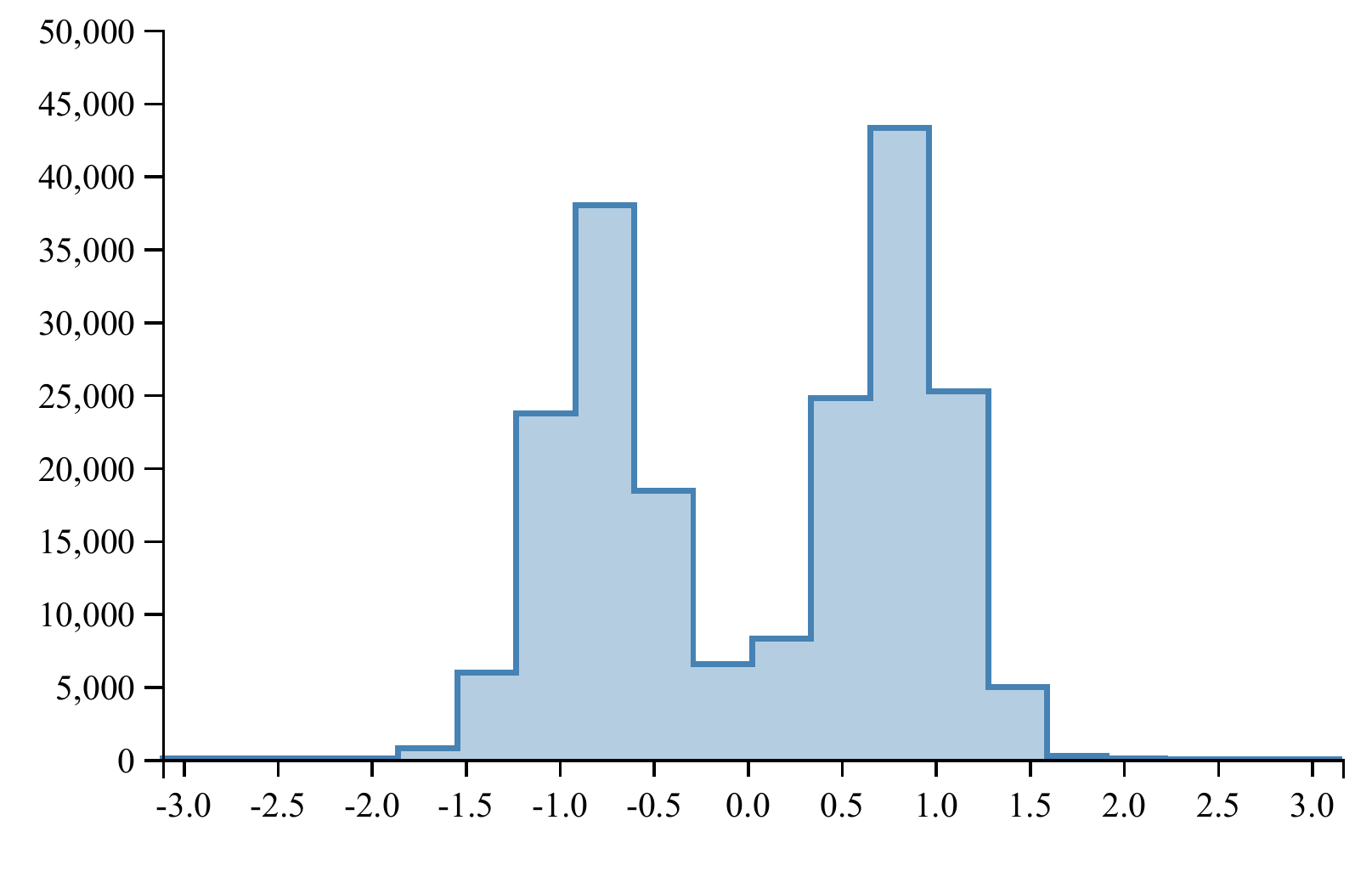}
\par\end{centering}
\begin{centering}
\includegraphics[width=0.3\textwidth]{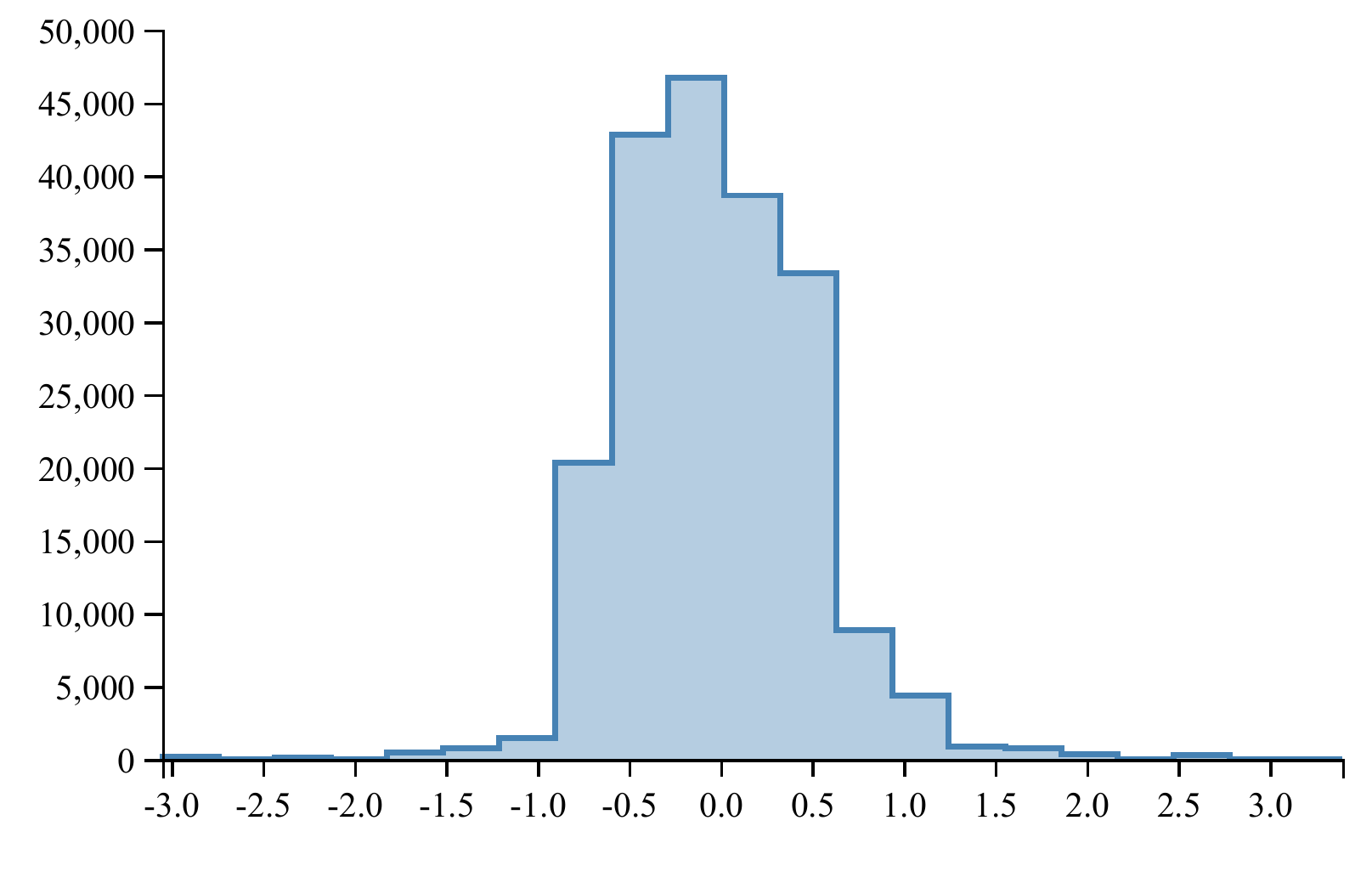}\includegraphics[width=0.3\textwidth]{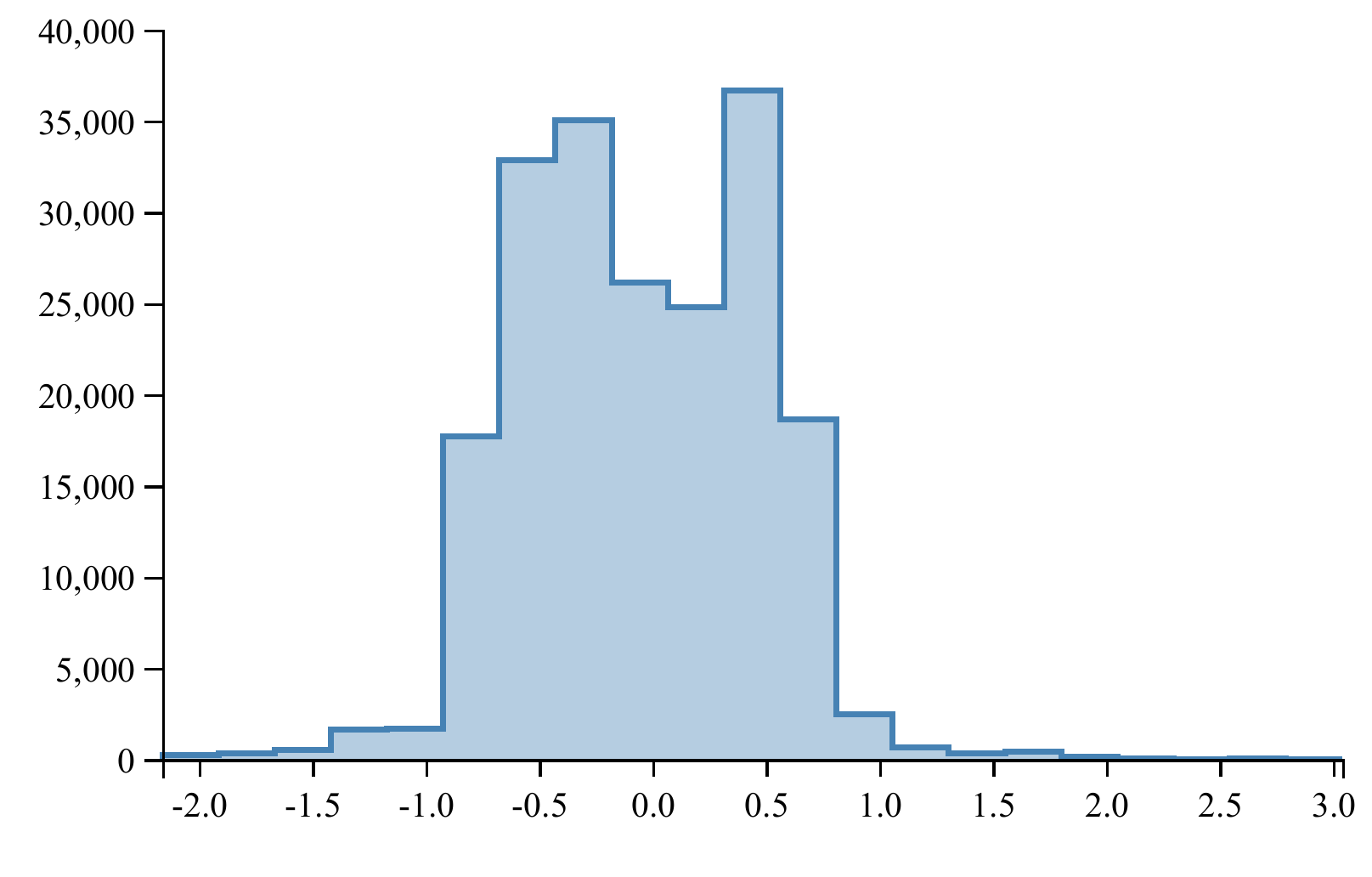}\includegraphics[width=0.3\textwidth]{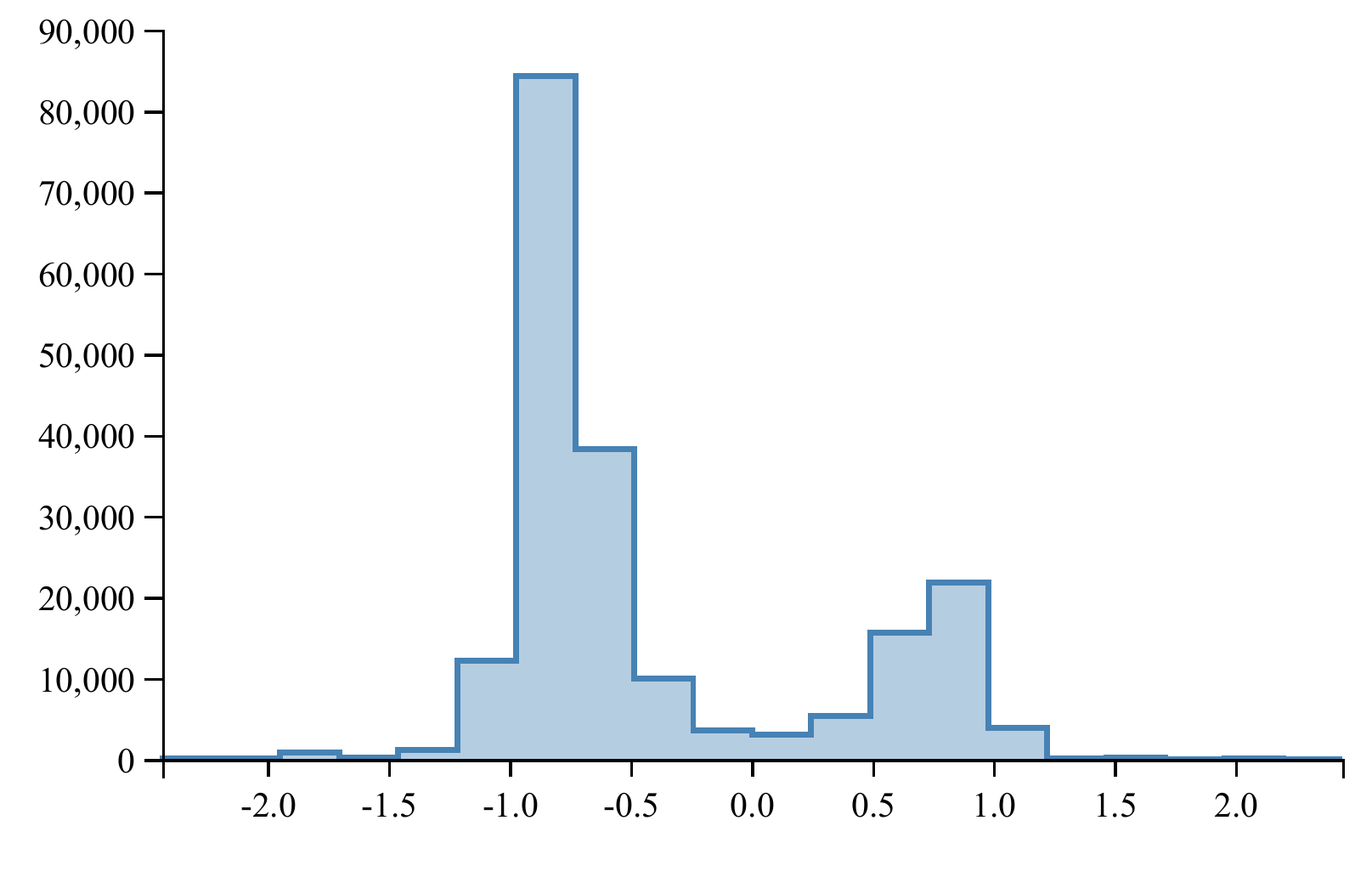}
\par\end{centering}
\begin{centering}
\includegraphics[width=0.3\textwidth]{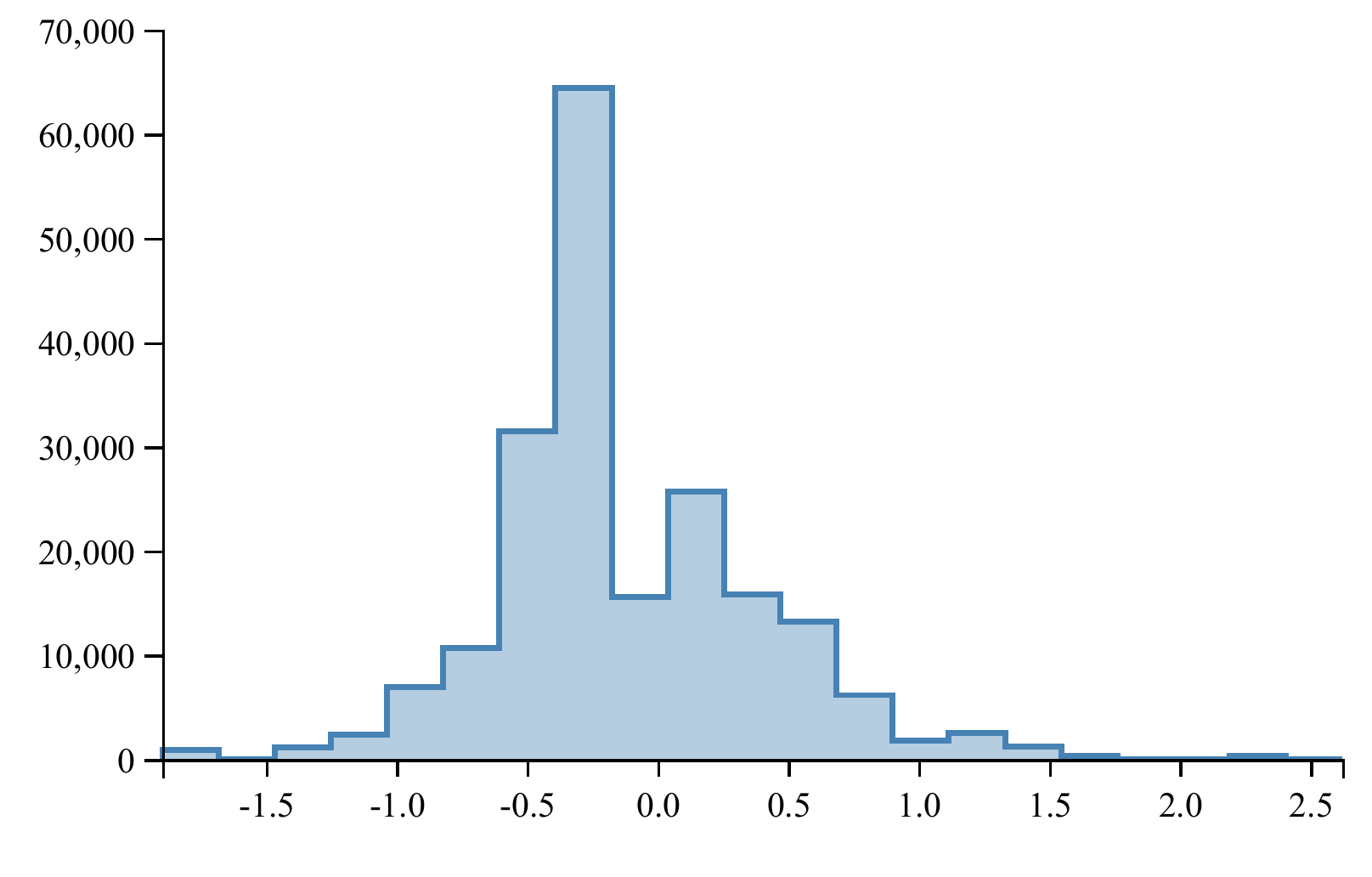}\includegraphics[width=0.3\textwidth]{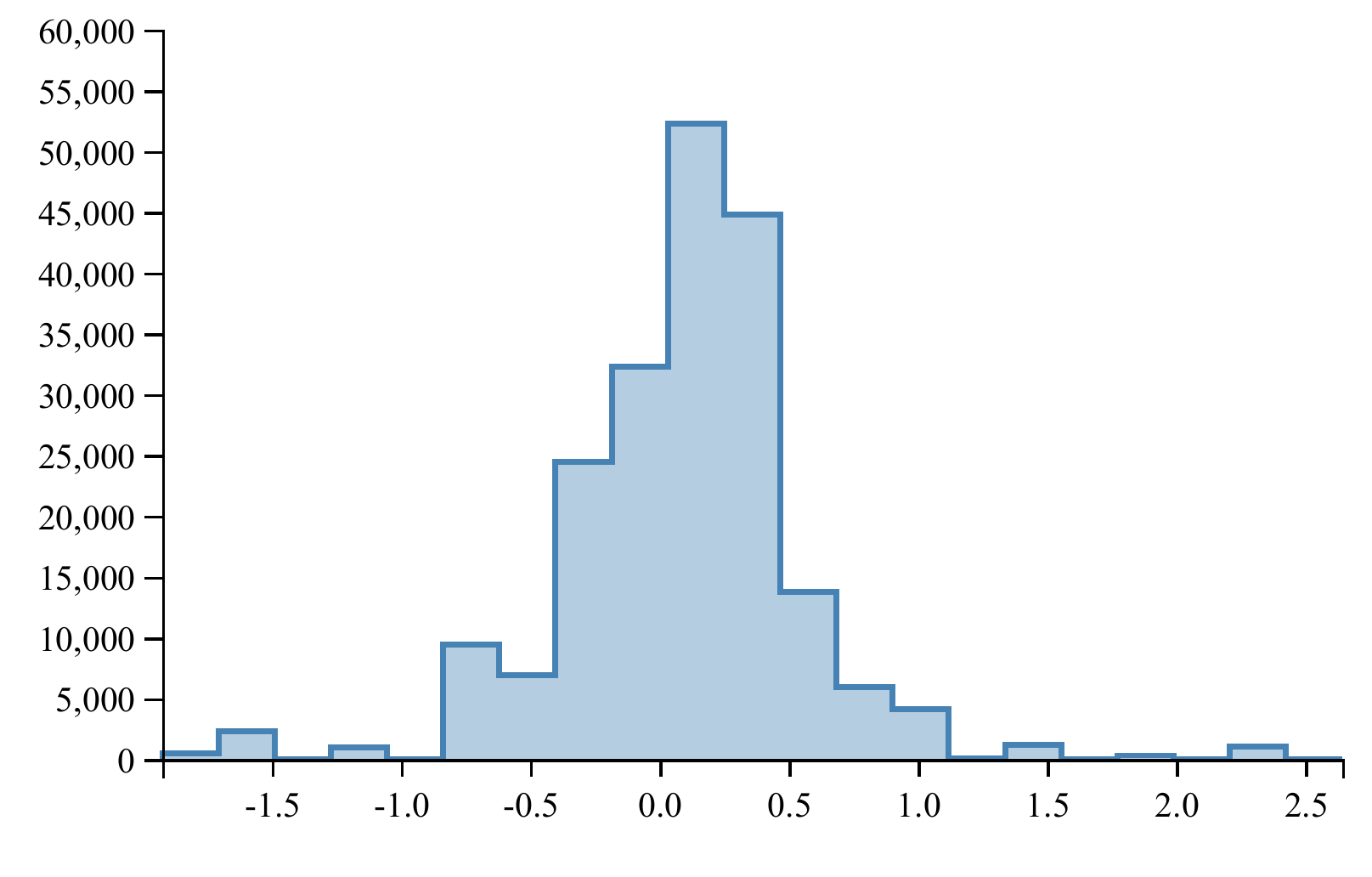}\includegraphics[width=0.3\textwidth]{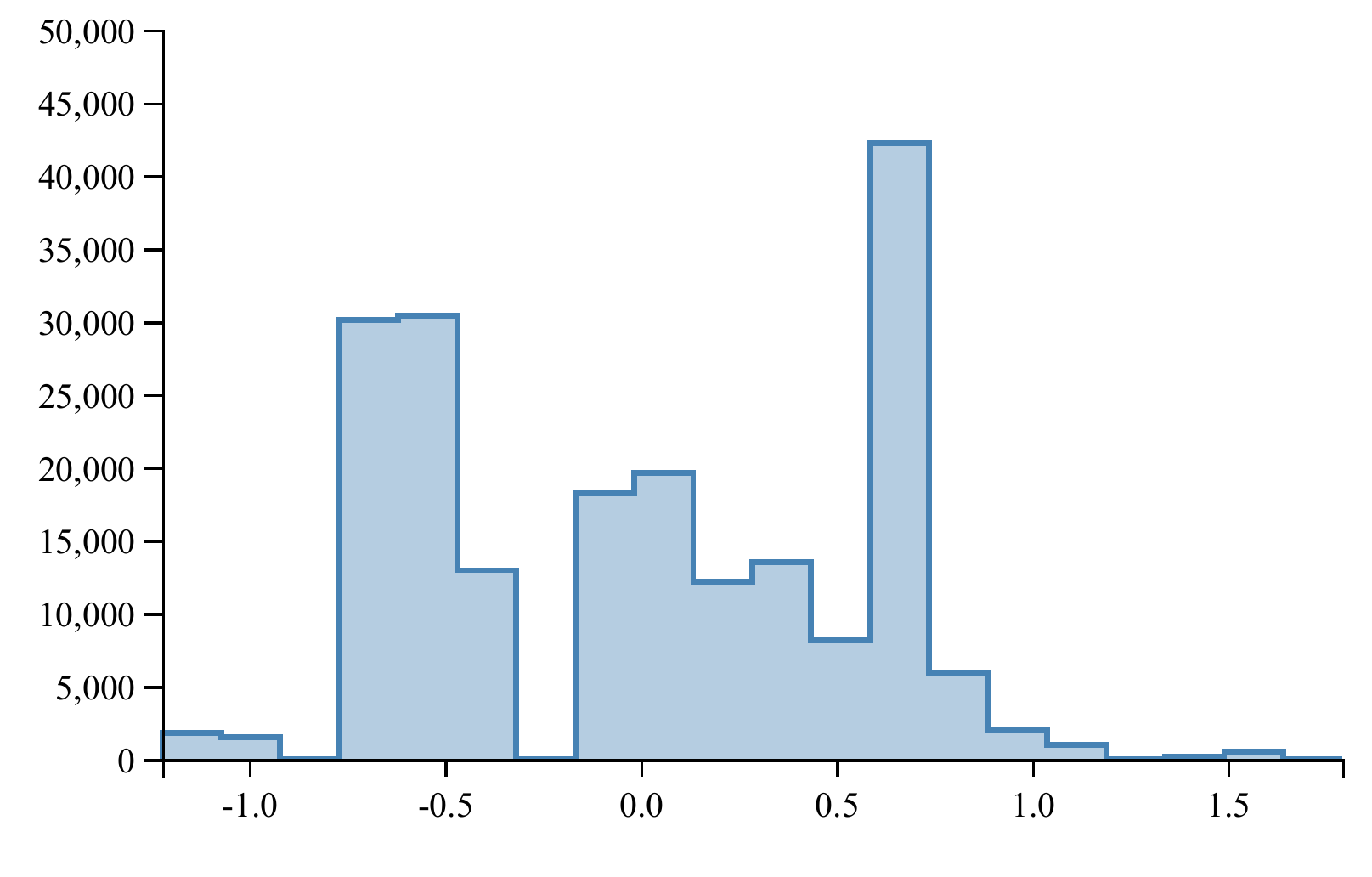}
\par\end{centering}
\caption{\label{fig:Political-Opinion-after-1}Political Opinion after observations.
From left: ME1, ME2 and ME3. From top to bottom: 1, 10 and 100 observations.}

\end{figure*}

Figures \ref{fig:Political-Opinion-after-1} shows $P_{a}$ distributions
after 1, 10 and 100 observations of news item for given media environments.

ME1 shows a Gaussian distribution of $p_{a}$ after prolonged exposure
to the environment. The strong majority of opinions are formed within
a moderate range of the political spectrum (i.e. within $\pm0.5$)
This is hardly a surprising result given the majority of the media
outlet in ME1 falls under such political spectrum.

ME2 is a more interesting case. Early distributions after 1 and 10
observations show a clear bi-modality of the distribution, which is
not surprising given the large portion of the partisan media. After
100 observations, however, the distribution looks a lot like ME1 distribution,
with a strong majority opinion formed within a moderate range. From
this result, it stands to reason that media taking editorial stances
on issues are healthy phenomena given that they are factual, and the
overall distribution of the partisan media is equally distributed
on both sides of the argument.

ME3 results show a high degree of polarization in the extreme range
of the political spectrum. After 10 observations, an alarming level
of concentration of the opinion is found on both extremes. This is
different from ME1 and ME2 results where you can still find a strong
majority within the moderate band. This shows the loss of common ground
in the public discourse in a Post-Truth politics. After 100 observations,
a moderate opinion does gain some of the shares back. However, the
majority is opinions are still outside of the moderate band, with
a very high concentration found on each of the extreme sides.

Overall, the result shows a detrimental effect that Fake News has
on public discourse according to our model. The result can account
for an unusually high level of opinion concentration on fringe opinion,
and an upsetting result of the public poll when the proposal is a
rather extreme one.

\section{Discussion and Conclusion}

The biggest limitation of this study is the lack of any data fitting
in designing the model. Although we did consider empirical results
from other studies for inspiration, the model itself is largely hand-wired
with our subjective view and understanding of the matter. In order
to tune our model to reflect the reality better, many of the magic
numbers appearing in our model need to be parameterized:
\begin{APAitemize}
\item prior distribution of $p_{a}$ and $a_{a}$
\item scale and decay rate in $d_{m}$ formulation
\item standard deviation in likelihood function
\end{APAitemize}
To fit those values, in the future study, we shall gather statistics
from existing surveys of the media environment, and design controlled
experiment to determine the parameters within the model. Finally,
another experiment shall be conducted to determine the accuracy of
the model compared to real-world experimental results.

Although the prescriptive analysis was not the main theme of this
study, the study can be further extended to account for such analysis.
The conclusion from studies with a prescriptive focus such as \citet{INNOCULATE}
and \citet{10.1371/journal.pbio.2002020} suggest that an antidote
to Fake News is the effective communication of scientific consensus.
Given the recurrent structure of our model, this can be translated
to observing moderate news items with high truth values prior to being
exposed to Fake News items.

In this study, we have essentially abstracted out news items as two
variables: politics and truthfulness. In the future study, it will
be interesting to give a more detailed account of the semantics of
news item. Given the recurrent structure of the model, it will be
natural to extend the model to interface with other types of Bayesian
recurrent neural network (\citet{fortunato2017bayesian}) that focuses
on natural language processing. Doing so will allow us to train the
model from primary sources of the data, namely the news items themselves.

What we have presented here is a way to model the behavior of individual
news-consuming agents. Given the advent of personal media and the
prevalence of social media, it will be interesting to devise a multi-agent
model (following \citet{Aymanns_2017}) where the interaction between
the agents are modeled after social networks, with varying degrees
of reaches and influences from multiple individuals.

The present study showcases an interesting use of probabilistic programming
to model complex social phenomena from the cognitive perspective of
individual agents. Future work ideas discussed in this section will
further the use of probabilistic programming and other computational
cognitive methods in the realms of computational social sciences.

\bibliographystyle{apacite}
\bibliography{citation}

\end{document}